\documentclass[twocolumn,showpacs,preprintnumbers,amsmath,amssymb]{revtex4}
\usepackage{amssymb}
\usepackage[dvips]{graphicx}
\begin{document}

\title{Magnetic quantum oscillations in nanowires}
\author{A. S. Alexandrov$^1$ and V. V. Kabanov$^{2}$}

\affiliation{$^1$Department of Physics, Loughborough University,
Loughborough, United Kingdom\\ $^{2}$Josef Stefan Institute 1001,
Ljubljana, Slovenia}

\begin{abstract}
 Analytical expressions for the magnetization and the longitudinal conductivity of
 nanowires are derived in a  magnetic field, $B$. We show that the interplay between
 size and magnetic field energy-level quantizations  manifests itself through novel
 magnetic quantum oscillations in metallic nanowires. There are three characteristic
 frequencies of  de Haas-van Alphen (dHvA) and Shubnikov-de Haas (SdH) oscillations,
 $F=F_0/(1+\gamma)^{3/2}$, and  $F^\pm=2F_0/|1+\gamma\pm (1+\gamma)^{1/2}|$,  in
 contrast with a single frequency $F_0= S_F\hbar c/(2\pi e)$ in simple bulk metals.
 The amplitude of oscillations is strongly enhanced in some "magic" magnetic fields.
 The wire cross-section area $S$ can be measured using  the oscillations as
 $S=4\pi^2 S_F\hbar^2c^2/(\gamma e^2B^2)$ along with the Fermi surface cross-section area,
 $S_F$.
\end{abstract}


\maketitle High magnetic fields have been widely used to explore
the single particle spectrum of bulk metals. Historically,  dHvA
and SdH quantum oscillations in magnetic fields have provided an
unambiguous signature and accurate quantitative information on the
Fermi surface and the damping of quasiparticles \cite{shon}.
Essential deviations from the conventional three-dimensional (3D)
oscillations have been found in low-dimensional metals like 2D
organic conductors \cite{sin,kar}. At present conducting nanowires
and nanotubes of almost any cross-section down to nanometer scale
and of any length can be prepared  with modern nano-technologies
\cite{aledem}. There are significant opportunities for discovery
of unique nanoscale phenomena arising from the dimension
quantization. In particular, galvanomagnetic transport properties
of nanowires have been the subject of many studies during last
decades \cite{her}.  Heremas et al. \cite{her} observed the
semimetal-semiconductor phase transition in the magnetoresistance
caused by the interplay between the electron cyclotron orbits, the
size energy-level quantization and the inter-band transfer of
carriers  in Bi nanowires. Their magneto-conductance was
theoretically addressed in the extreme 1D limit \cite{sinmol}. The
Aharonov-Bohm-type  oscillations of the magneto-conductance have
been discovered in carbon nanotubes \cite{bac,fuj} and connected
with a metal-insulator transition caused by shifting of the van
Hove singularities of the density of states \cite{roc}. More
recently SdH oscillations were observed in arrays of 80 nm
Bi-nanowires \cite{hub} and in 200nm Bi-nanowires \cite{gro} in
first and second derivatives of resistance with respect to the
magnetic field. There is a great demand for quantitative
characterization of nanowires and  analytical descriptions of the
interplay between dimension and field-induced energy-level
quantizations.

In this Letter, we present the theory of magnetic quantum
oscillations  in long metallic nanowires in the longitudinal
magnetic field, ${\bf B}$, parallel to the direction of the wire
$z$.  We consider clean nanowires with the electron mean free
path, $l=v_F \tau$  comparable or larger than the cross size, $R$,
but smaller than the nanowire length, $L$, which allows us to
apply the conventional Boltzmann kinetics.   We also assume that
the electron wavelength near the Fermi level is very small in the
\emph{metallic} nanowires, so that $L\gg l\gtrsim R \gg 2\pi \hbar
/(m^*v_F)$, where $v_{F}$ is the Fermi velocity and $m^*$ is the
band mass in the bulk metal. We find novel quantum oscillations of
the magnetization and the conductivity caused by the interplay
between  magnetic  and  dimension energy-level quantizations.

Let us first calculate the magnetization
$M=-\partial\Omega/\partial B$, where $\Omega =-k_BT \sum_{\alpha}
\ln [1+\exp(-\xi_{\alpha}/(k_BT))]$ is the thermodynamic potential,
$\xi_{\alpha}=E_{\alpha} -\mu$, $E_{\alpha}$  is the
single-particle energy spectrum and $\mu$ is the chemical
potential.

Boundary conditions on the surface of the wire are not compatible
with the symmetry of the vector potential, ${\bf A}= {\bf B}
\times {\bf r}/2 $, so there are no simple analytical solution for
$E_{\alpha}$ in the magnetic field. However, one can overcome this
difficulty in the quasi-classical limit, $\mu\gg \hbar \omega_s$, where
$\omega_s\equiv \pi v_F/R$,  using the Tomonaga-like linearization
of the energy spectrum \cite{tom}. Approximating the wire as an
infinite round well one obtains $E_{\alpha}=
\hbar^2(k_{nm}^2+k^2)/(2m^*)$. Here $\hbar k$ is the continuous momentum along
the wire, and discrete $k_{nm}$ are defined as zeros of the Bessel
functions $J_{|m|}(k_{nm}R)=0$, where $m=0,\pm 1,\pm2, ...$ are
the eigenvalues of z-component of the orbital momentum. In the
quasi-classical limit $J_{|m|}(k_{nm}R)\approx (k_{nm}R)^{-1/2}
\cos (k_{nm}R -\pi |m|/2 - \pi/4)$, and $k_{nm}R=\pi (2n+1)/2 +\pi
|m|/2 + \pi/4 $ with $n=0,1,2,...$.

Hence,  near the Fermi surface the spectrum is given by
$\xi_{n,m,k}\approx \hbar \omega_s (2n + |m|-n_F) +\hbar^2 k^2/(2m^*)$, which is
identical to the spectrum in a parabolic "confinement" potential
$V({\bf r})= m^*\omega_s^2 (x^2+y^2)/2$ (here $n_F= \mu/(\hbar \omega_s)
\gg 1$). The major contribution to dHvA and SdH oscillations
arises from the energy spectrum near the Fermi level, so we can
replace the metallic nanowire with the confinement potential. In
contrast with the original problem, the model Hamiltonian,
$H=({\bf p}- e{\bf A}/c)^2/(2m^*) + V({\bf r})+s\mu_BB$ has simple
analytical eigenfunctions, $\psi_{\alpha}({\bf r})\propto
\exp(ikz)\rho ^{|m|}\exp(-\rho^2/2) L_{n}^{|m|} (\rho^2)$ and
eigenvalues
\begin{equation}
E_{\alpha}={\hbar^2 k^2\over{2m^*}} +2 \hbar \omega\left(n+ {|m|-m+1\over{2}}
\right) + m \hbar \omega^- + s\mu_BB,
\end{equation}
where $\rho^2=(m^* \omega/\hbar)(x^2+y^2)$, $\omega^2=
\omega_s^2+\omega_c^2/4$, $\omega^\pm=\omega\pm \omega_c/2$,
$\omega_c=eB/m^*c$, $\mu_B$ is the Bohr magneton, and $\alpha =
\{n,m,k,s\}$ comprises all quantum numbers including the spin
$s=\pm 1$.
Using Eq.(1) and
replacing negative $m$ with $-m-1$ one obtains
 \begin{equation}
\Omega =-k_BT L \sum_{s,\pm}\int{dk\over {2\pi}} \sum_{n, m\geqslant 0}
\ln \left[1+\exp\left({\mu_s^\pm -\epsilon_{nm}^\pm(k)\over{k_BT}}
\right)\right],
\end{equation}
where $\mu_s^+= \mu -\hbar \omega -\mu_B Bs$, $\mu_s^-= \mu -\hbar (\omega
+\omega^+) -\mu_B Bs$, and $\epsilon_{nm}^\pm(k)=2\hbar n \omega
+\hbar m \omega^\pm +\hbar^2 k^2/(2m^*)$. Summations over $n$ and $m$ can be
replaced by  sums over $r,r'= 0,\pm1,\pm2,....\pm\infty$ using
twice the Poisson's formula and  the variables $x=2\omega n
+\omega^\pm m$ and $y=\omega n-\omega^\pm m/2$ in place of $ n$
and $ m$,
\begin{eqnarray}
&&\sum_{n, m\geqslant 0} f(2\omega n +\omega^\pm m)=
\sum_{r,r'}{1\over{2\pi i(r\omega^\pm-2\omega r')}} \times \cr
&&\int_0^\infty dx f(x) \left[\exp\left({2\pi i
rx\over{2\omega}}\right) -\exp\left({2\pi i
r'x\over{\omega^\pm}}\right)\right].
\end{eqnarray}
We are interested  in an oscillatory correction, $\tilde{\Omega}$
to the thermodynamic potential arising from the terms in Eq.(3)
with nonzero $r$ or $r'$. Introducing a new variable
$\xi=x+\hbar^2 k^2/(2m^*)-\mu_s^\pm $, integrating by parts, extending the
lower limit of $\xi$ down to $-\infty$ and taking routine
integrals over $k$, $\int dk \exp(iak^2)= (\pi/|a|)^{1/2}
\exp[i\pi a/(4|a|)]$ and over $y=\xi/(k_BT)$, $\int dy \exp(iay)
[1+\exp(y)]^{-1}=-i\pi/\sinh(\pi a)$, we finally obtain
\begin{eqnarray}
&&\tilde{\Omega}= \sum_{r=1}^{\infty} \sum_{\pm}
A_r (\omega, \omega^\pm) \sin\left({\pi r \mu\over{\hbar \omega}}
-{\pi r (\omega^+\mp \omega^-)\over{2\omega}}-{\pi\over{4}}\right) \cr
 &+& A_r(\omega^\pm/2, 2\omega) \sin\left({2\pi r \mu\over{\hbar \omega^\pm}}
\pm{\pi r (\omega^+\mp \omega^-)\over{\omega^\pm}}-{\pi\over{4}}\right),
\end{eqnarray}
where
\begin {equation}
A_r(x,y)= {k_B T L(2m^*x/\hbar)^{1/2} \cos [\pi r\mu_B B/(\hbar x)]\over{2\pi r^{3/2}
\sinh[\pi^2k_BTr/(\hbar x)]}}
\cot \left({\pi r y\over{2x}}\right)
\end{equation}
are oscillation amplitudes, and the summation formula
$\sum_{r} (z-r)^{-1}= \pi \cot (\pi z)$ has been
applied.

Here and further we neglect quantum oscillations of the chemical
potential. For the sake of transparency, we also neglect a damping
of quantum levels by the impurity scattering in dHvA oscillations.
We introduce this damping in the SdH effect (see below) neglecting
quantum oscillations of the scattering rate $1/\tau$. The quantum
oscillations of $\mu$ and $1/\tau$ could lead to a mixing of dHvA
frequencies in multi-band metals as predicted and experimentally
observed in  several bulk compounds \cite{alebra,sin2,shep,yos}.
However, they are negligible in the presence of a field and
size-independent "reservoir" of states (i.e. a sub-band with a
heavy mass \cite{alebra2}) and the inter-band scattering.

There are three characteristic frequencies $2\omega$ and
$\omega^\pm$ in the oscillating part of the magnetization
$\cal{M}$ $= -\partial \tilde{\Omega}/\partial B$,  rather then a
single frequency $\omega_c$.
\begin{figure}
\begin{center}
\includegraphics[angle=-0,width=0.47\textwidth]{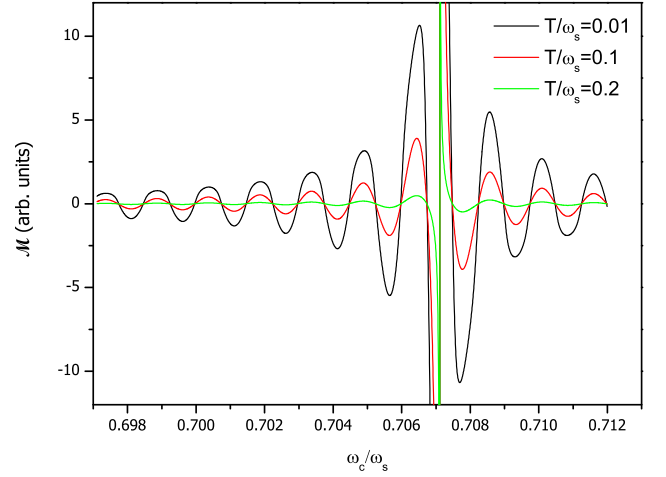}
\vskip -0.5mm \caption{Oscillating part of the magnetization
versus the magnetic field for relatively low fields and three
temperatures. The  resonance at $\omega_c=\omega_s/\sqrt{2}$ is
due to a partial recovery of the  energy-level degeneracy.}
\end{center}
\end{figure}

The same frequencies are found in the  conductance, $\sigma$.  The
longitudinal conductivity is given by \cite{agd}
\begin{eqnarray}
&&\sigma({\bf r},\nu_0)={ie^2 \hbar^2 k_BT\over{2\nu_0 (m^*)^2}}\sum_{s,\omega_p}
\left({\partial \over{\partial z}}-{\partial\over{\partial z'}}
\right)_{{\bf r'}\rightarrow {\bf r}} \times \cr
&&\int d{\bf r''}\overline{ G_s({\bf r}, {\bf r''};
\omega_p){\partial\over{\partial z''}}G_s({\bf r''},
{\bf r'}; \omega_p-\nu_0)} \cr
&-&{ie^2 k_B T\over{\nu_0 m^*}}\sum_{s,\omega_p}\overline{ G_s({\bf r},
{\bf r}; \omega_p)}, \nonumber
\end{eqnarray}
where $\hbar \omega_p=\pi k_BT (2p+1)$, $p=0, \pm 1, \pm 2, ...$, and
$\nu_0=2\pi n k_BT/\hbar$  is the "frequency" of the  "time"-dependent
z-component of the vector potential, $A_z= iEc\nu_0^{-1}
\exp(-i\nu_0 t)$, due to a longitudinal electric field $E$(
$0\leqslant t \leqslant \hbar/(k_BT)$).  The static conductivity is
calculated as the analytical continuation of this equation to
$\nu=i\nu_0 \rightarrow 0$. The product of two GFs  averaged over
the random impurity distribution is factorized as the product of
averaged GFs for a short-range scattering potential in absence of
vertex corrections \cite{agd},
\begin{equation}
\overline{ G_s({\bf r}, {\bf r'}; \omega_p)}= \sum_{n,m,k}{\psi_{\alpha}({\bf r})
\psi^*_{\alpha}({\bf r'})\over{i\hbar\tilde{\omega_p} -\xi_{\alpha}}}, \nonumber
\end{equation}
where $\tilde{\omega_p}=\omega_p+\omega_p/(2|\omega_p|\tau)$.
Then integrating the conductivity $\sigma({\bf r}, \nu_0)$ over
the cross-section of the wire one obtains the conductance,
\begin{eqnarray}
\sigma(\nu)&=&-{ie^2 k_B T\over{L \nu_0 m^*}}\sum_{\alpha,\omega_p}
 {\hbar^2 k^2/m^* \over{(i\hbar \tilde{\omega_p} -\xi_{\alpha})[i\hbar (\tilde{\omega_p}-\nu_0)
-\xi_{\alpha}]}}\cr
&+&{1\over{i \hbar \tilde{\omega_p} -\xi_{\alpha}}}.
\end{eqnarray}
Integrating by parts the second diamagnetic term in Eq.(6)
cancels the paramagnetic part at $\nu_0=0$. The routine analytical
continuation \cite{lif} of  the remaining paramagnetic part
yields the static conductance \cite{ref} in the limit $\nu
\rightarrow 0$,
\begin{equation}
\sigma=-{e^2 \hbar^3 \over{L \pi (m^*)^2}} \sum_{\alpha} k^2
\int_{-\infty}^{\infty} d\xi {\partial f(\xi)
\over{\partial \xi}} \left[\Im G^{R}_{\alpha}(\xi)\right]^2,
\end{equation}
where $ G^{R}_{\alpha}(\xi)= [\xi-\xi_{\alpha}-i\hbar/(2\tau)]^{-1}$
is the retarded GF and $f(\xi) =[1+\exp \xi/(k_BT)]^{-1}$.

Summations over $n$ and $m$ are performed  using twice the
Poisson's formula, as in Eq.(3). The term with $r=r'=0$ yields the
classical contribution,
\begin{eqnarray}
&&\sigma_0=\sum_{s,\pm}{e^2  \tau \over{4\hbar^3 \pi\omega
\omega^\pm (m^*)^{1/2}}} \int_{0}^{\infty} dx x \times \cr
&&{[\mu^\pm_s -x + ((\mu^\pm_s -x )^2+\hbar^2/(4\tau^2))^{1/2}]^{3/2}
\over{((\mu^\pm_s -x )^2+\hbar^2/(4\tau^2))^{1/2}}}
\end{eqnarray}
after integrating over $k$ and neglecting temperature corrections.
One can also neglect $\hbar^2/(4 \tau^2)$ in the integral, Eq.(8) and
obtain the conventional Drude conductance, $\sigma_0= Ne^2
\tau/(Lm^*)$, where $N=8 L (2m^*)^{1/2} \mu^{5/2}/ (15\hbar^3 \pi \omega_s^2)$
is the total number of electrons in the wire. Calculating  quantum
corrections in $\sigma=\sigma_0+\tilde{\sigma}$ is similar to
calculating of $\tilde{\Omega}$. Using the integrals $\int dk
k^2\exp(iak^2)= (i\pi^{1/2}/2)/|a|)^{3/2} \exp[i\pi a/(4|a|)]$ and
$\int dy \exp(iay) \cosh^{-2}(y)=-\pi a/\sinh(\pi a/2)$  we
obtain
\begin{eqnarray}
&&\tilde{\sigma}= \sum_{r=1}^{\infty} \sum_{\pm} B_r
(\omega, \omega^\pm) \cos\left({\pi r \mu\over{\omega}}
-{\pi r (\omega^+\mp \omega^-)\over{2\omega}}-{\pi\over{4}}\right) \cr
 &+& B_r(\omega^\pm/2, 2\omega) \cos\left({2\pi r \mu\over{\omega^\pm}}
\pm{\pi r (\omega^+\mp \omega^-)\over{\omega^\pm}}-{\pi\over{4}}\right),
\end{eqnarray}
 where
\begin{eqnarray}
B_r(x,y)&=& {e^2 \tau k_B T \cos [\pi r\mu_B B/(\hbar x)] \exp[-\pi r/(2x\tau)]
\over{\hbar(2m^*\hbar x)^{1/2} r^{1/2} \sinh[\pi^2 k_B Tr/(\hbar x)]}}\cr
&\times& \cot \left({\pi r y\over{2x}}\right).
\end{eqnarray}
\begin{figure}
\begin{center}
\includegraphics[angle=-0,width=0.47\textwidth]{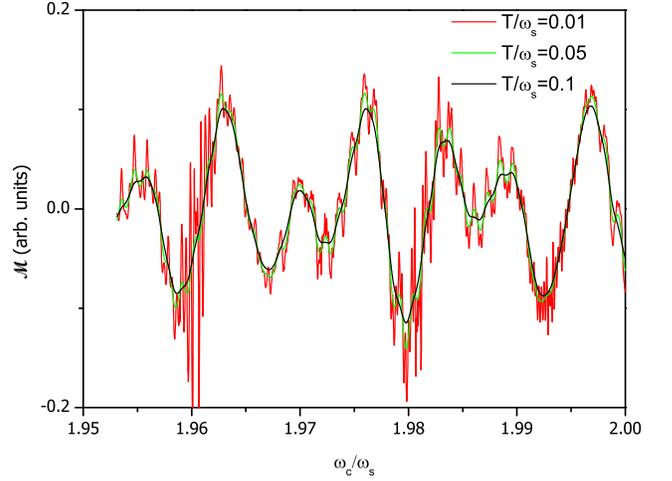}
\vskip -0.5mm \caption{Oscillating  magnetization for intermediate
fields and three temperatures. Magic resonances are observed in
many Fourier harmonics at low temperatures. }
\end{center}
\end{figure}
If the conventional dHvA frequency is high, $F_0
\gg B$,  three novel dHvA/SdH frequencies, $F, F^\pm$ of
the wire can be estimated as $F=B^2/\delta B \simeq \mu B^2 |df/dB|/(\hbar f^2)$
with $f=2\omega, \omega^\pm$,
\begin{equation}
F=F_0/(1+\gamma)^{3/2},
\end{equation}
and
\begin{equation}
F^\pm=2F_0/|1+\gamma\pm (1+\gamma)^{1/2}|,
\end{equation}
where $F_0= S_F \hbar c/(2\pi e)$, $\gamma=4\omega_s^2/\omega_c^2=
4\pi^2 S_F\hbar^2c^2/(e^2SB^2)$, $S=\pi R^2$ is the cross-section
area of the wire, and $S_F=\pi (m^*)^2v_F^2/\hbar^2$ is the Fermi-surface
cross-section area. They are related as $F=
F^+F^-|F^+-F^-|/(F^++F^-)^2$.

\begin{figure}
\begin{center}
\includegraphics[angle=-0,width=0.47\textwidth]{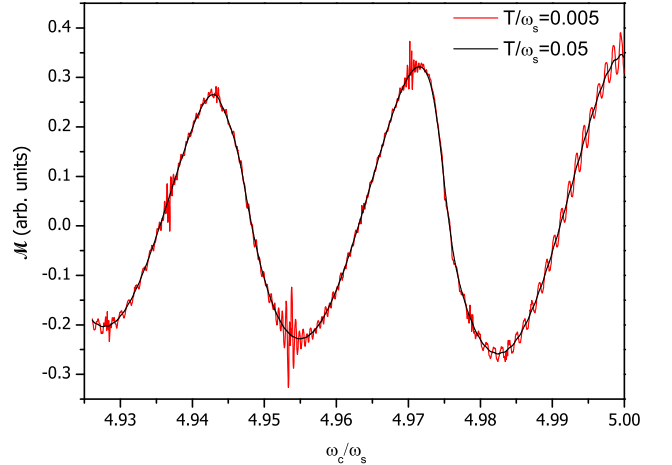}
\vskip -0.5mm \caption{Oscillating  magnetization  for high fields
and two  temperatures.}
\end{center}
\end{figure}
Remarkably, both temperature and scattering damping factors  in
Eqs.(5,10) depend on $\omega$ and $\omega^\pm$ rather than on the
cyclotron frequency $\omega_c$. Hence there are no constraint on
the value of the magnetic field imposed by those factors as soon
as $\omega_s$ is large enough, $\omega_s > T, 1/\tau$. In low
fields, where $\gamma \gg 1$, all  frequencies are much lower than
$F_0$, $F\approx F_0/\gamma^{3/2}$ and $F^\pm\approx 2F_0/\gamma$.
In high fields, where $\gamma \ll 1$, two of them are about the
same as $F_0$, $F\approx F^+ \approx F_0$, while the third one
appears to be much higher, $F^-\approx 4F_0/\gamma\gg F_0$.  With
respect  to the Pauli paramagnetism and Landau diamagnetism in the
bulk metal, amplitudes of  quantum corrections in the
magnetization and in the magnetic susceptibility, $\chi$, per unit
volume  are about $(\hbar\omega_s/\mu)^{1/2}$ and
$[\mu/(\hbar\omega_s)]^{1/2}$, respectively,  (to get these estimates we
divide $\cal{M}$ by $\pi R^2$). The relative amplitude of quantum
corrections in the conductance is    $(\hbar \omega_s/\mu)^{5/2}$, and
about $(\hbar \omega_s/\mu)^{3/2}$ and $(\hbar \omega_s/\mu)^{1/2}$ in its
first and second field derivatives, respectively.
If we take $\omega_s$ about the same as $\omega_c$, the quantum corrections are
much smaller than in the bulk metal,
where they have the relative order of magnitude as
$[\mu/(\hbar \omega_c)]^{1/2}$ in $\cal{M}$,  $[\mu/(\hbar\omega_c)]^{3/2}$ in
$\chi$ and  $(\hbar \omega_c/\mu)^{1/2}$ in $\sigma$ \cite{shon}.
However, there are some "magic" magnetic fields where the quantum corrections
"explode". These are fields where  the condition
$2\omega/(\omega^\pm) =(q+2)/r$ is satisfied, so   $"\cot"$ in
Eqs.(5,10) becomes infinite if $q$ is an integer. In particular,
first harmonics with $r=1$ become infinite if
\begin{equation}
{\omega_c\over{\omega_s}}= {q\over{\sqrt{q+1}}}={1\over{\sqrt{2}}},
{2\over{\sqrt{3}}}, {3\over{\sqrt{4}}},....
\end{equation}
These magic resonances are clearly seen in Figs.1,2
at low temperatures, where we present numerical data for the
oscillating part of the magnetization ($\mu/\omega_s$ is $1000$
and  we choose $\cos [\pi r\mu_B
B/(\hbar x)]=1$).
At high fields, $\omega_c \gg \omega_s$, the conventional dHvA
pattern dominates,  but the magic resonances are still there,
Fig.3.

Let us elaborate more about the physical origin of the magic resonances.
It is well known that the Landau levels are $ SeB/(2\pi c \hbar)$-fold degenerate
in the bulk metal of the cross-section area $S$. The boundary conditions in the
nanowire (approximated here by the confinement potential) remove the degeneracy,
Eq.(1). Therefore the density of states at every level is reduced by a factor
$\omega_s^2/(\pi^2 \mu \omega_c)$, which explains the reduction of quantum
amplitudes compared with the bulk metal. However,  the magic resonance conditions
partially restore  the degeneracy of the spectrum, Eq.(1). For example, if
$\omega_c=\omega_s/\sqrt{2}$, one obtains $2\omega= 3\omega_s/\sqrt{2}$ and $
\omega^-= \omega_s/\sqrt{2}$, so that
$E_{\alpha}=\hbar^2 k^2/(2m^*) + \hbar \omega_s(6n+ 3|m|-m+3)/(2\sqrt{2})
 + s\mu_BB$, which is the same for all combinations of $n$ and $m$ with a
 fixed value of $6n+ 3|m|-m$.
Hence, compared with the  amplitudes estimated above, the magic
amplitudes are enhanced.  The "anharmonic" corrections to the linearised
energy spectrum in Eq.(1)imposed by the boundary conditions restrict their enhancement.

It might be
difficult to observe the novel  oscillations in the magnetization
of a single nanowire because its  small volume, but they could be
measured on bundles of nanowires. As far as SdH oscillations
in nanowires \cite{hub,gro} is concerned, their quantitative
comparison with the present theory needs  measurements  in a wider
field-range allowing for the reliable Fourier analysis.
Using the typical radius of $Bi$-nanowires  $R=100$ nm \cite{sinmol,hub,gro} and the
Fermi surface cross-section area  $S_F=10^{13} cm^{-2}$  \cite{bhar} yields an
estimate of $\hbar \omega_s/k_B \approx 50 K$ with the carrier mass $m^*=0.1 m_e$.
Then the lowest temperature presented in Figs. 1, 2  is about $0.5$ K with these parameters.

In conclusion, we have presented the  theory  of magnetic quantum
oscillations in clean metallic nanowires with  simple
Fermi-surfaces. We have found novel  oscillations caused by the
interplay between  size and field energy-level quantizations with
three characteristic frequencies, calculated their amplitudes and
identified magic resonances, where the quantum corrections are
strongly enhanced. Our findings suggest that one can measure both
reciprocal and real space geometries of nanowires in a single
measurement.

\end{document}